\documentstyle[12pt]{article}
\begin{document}
\begin{center}
\def\thefootnote{*}
{\large {\bf ONCE MORE ON THE RADIATIVE CORRECTIONS TO THE NUCLEON
STRUCTURE FUNCTIONS IN QCD}}\footnote{This work has been supported
by Bulgarian Science Foundation under Contract F 16}\\
\vskip 1truecm
\def\thefootnote{**}
{\bf D.B. Stamenov}\footnote{Permanent address:Institute for Nuclear
Research and Nuclear Energy,
Bulgarian Academy of Sciences, Boul. Tsarigradsko chaussee
72, Sofia 1784, {\bf Bulgaria}\\
Telefax:~(359)-2-75-50-19, ~E-mail: stamenov@bgearn.bitnet}\\
\normalsize International Centre for Theoretical Physics\\
P.O. Box 586, Trieste 34100, {\bf Italy}\\[2mm]
\vskip 2truecm
{\bf Abstract} \\[1mm]
\end{center}

    A new representation of the next to leading QCD corrections to
the nucleon structure functions is given in terms of parton
distributions. All $~O(\alpha_{s})~$ corrections to the leading
logarithmic approximation (LLA) are included. In contrast to the similar
representations in the literature terms of order $~O(\alpha_{s}^{2})~$ do
not attend in our expressions for the nucleon structure functions
taken in the next to leading logarithmic approximation. This result
is generelized for any order in $\alpha_{s}$ beyond the
LLA. Terms of order $~O(\alpha_{s}^{n})~$ which belong only to the
approximation in consideration are present in such a representation
for the structure functions.\\[1.2cm]

\newpage \begin{flushleft}
{\large {\bf 1. Introduction}}
\end{flushleft}
\vspace*{1mm}

     To make quantitative predictions for hadron processes at
present and future high-energy colliders detailed information on
the parton distributions inside the nucleon is needed. The study of
deep inelastic lepton-nucleon processes is one of the best manners
to obtain this information. The increased precision of the deep
inelastic scattering data enables us to derive the quark and gluon
distributions in the framework of the QCD improved parton model
with a good degree of accuracy.\\

     The nucleon structure functions can be written in QCD as
suitable convolutions of quark, antiquark and gluon distributions.
At the leading logarithmic approximation (LLA) the results of the
QCD corrections can be interpreted by saying that the nucleon
structure functions $F_{i}(x,Q^{2})$ are given by the naive
parton model formulae expressed in terms of $Q^{2}$ dependent
effective parton distributions obeying to the first order
Lipatov-Altarelli-Parisi equations [1].\\

     Beyond the LLA the explicit form of the $\alpha_{s}$ corrections to
the structure functions depends on the definition of the effective
parton distributions. Different definitions of the parton distributions
will give, of course, the same physical results. In general, there
exist two different ways to define these distributions:\\

     i) A-definition [2,3] or so-called DIS scheme: All of the next
to LLA $\alpha_{s}$ corrections to one of the structure functions, for
instance $F_{2}$, are incorporated in the quark and antiquark
distributions. According to such a definition the parton model
formula for $F_{2}(x,Q^{2})$ is valid to all oders in perturbation
theory. In this case the form of the $\alpha_{s}$ corrections to the
other structure functions is very simple.\\

    ii) B-definition [4,5]: According to this definition of the
parton distributions the $\alpha_{s}$ corrections coming from the
Wilson coefficient functions are explicitly factored out. The latter are
structure functions dependent.\\

 Notice that in the framework of each of these definitions of
the parton distributions different renormalization-prescription
schemes can be used for the calculation of the radiative
corrections to the structure functions.\\

     The expressions for the nucleon structure functions presented
in the literature in terms of parton distributions have a
following common feature (independent on the definition). In the
next to leading logarithmic approximation (NLLA) of QCD for the
structure functions (all $O(\alpha_{s})$ corrections to LLA are taken
into account) there exist some terms of order $O(\alpha_{s}^{2})$ which do
not belong to this approximation. In any case, the presence of these
terms which are only a small part of $O(\alpha_{s}^{2})$ corrections is
undesirable. First of all, the moments of these structure functions do
not coincide with the moments calculated in NLLA in the formal
Quantum Field Theory (QFT) approach. Moreover, the quantity of the
$O(\alpha_{s}^{2})$ corrections may be changed considerably if all
corrections in this order are taken into account. The calculations of the
$O(\alpha_{s}^{2})$ corrections are in progress [6 - 8], but unfortunately,
some of them are still unknown.\\

     In this paper we give expressions for the next to leading
logarithmic approximation of the nucleon structure functions written in
terms of parton distributions in which the unwanted terms of order
$O(\alpha_{s}^{2})$ do not appear in this approximation. This
representation is generelized also for any order in $\alpha_{s}$ beyond
the LLA. There is one-to-one correspondence between such a representation
of the structure functions and the structure functions calculated in the
formal QFT approach.\\

\begin{flushleft}
{\large {\bf 2. Results}}
\end{flushleft}
\vspace*{1mm}

    In order to simplify the problem we shall restrict our discussion to
the nonsinglet part of the structure functions. At the end of the
paper we shall present also our results for the complete electromagnetic
and neutrino (antineutrino) structure functions.\\

     In the formal QCD approach the $Q^{2}$ dependence of the moments
of deep inelastic nonsinglet structure functions is given as

\begin{eqnarray*}
M_{i}^{NS}(n,Q^{2})&=&\int _{0}^{1} dx x^{n-2} {\cal F}_{i}^{NS}(x,Q^{2})\\
&=& A_{n}^{NS}(Q_{0}^{2})exp\{-\int _{\alpha_{s}(Q_{0}^{2})}
^{\alpha_{s}(Q^{2})}da{\gamma_{n}^{NS}(a)\over \beta(a)}\}C_{i,n}^{NS}
(1,\alpha_{s}(Q^{2})) .\begin{flushright}(1)\end{flushright}
{}~~~~~~~~~~~~(i=1,2,3)\end{eqnarray*}

The functions ${\cal F}^{NS}_{i}(x,Q^{2})$ are related to the
standard deep inelastic structure functions $F_{i}^{NS}(x,Q^{2})$ by

$${\cal F}^{NS}_{1}(x,Q^{2}) = 2xF^{NS}_{1}(x,Q^{2}),~~~~
{\cal F}^{NS}_{2}(x,Q^{2}) = F^{NS}_{2}(x,Q^{2}),$$
$${\cal F}^{NS}_{3}(x,Q^{2}) = xF^{NS}_{3}(x,Q^{2}).\eqno{(2)}$$

In (1) $A_{n}^{NS}(Q^{2}_{0})$ stand for the unknown hadronic matrix
elements of spin-$n$ nonsinglet operators and $C^{NS}_{i,n}$ are the
coefficient functions of these operators in the Wilson expansion [9].
$\alpha_{s}, ~\gamma^{NS}_{n}$ and $\beta$ are the the well known
effective coupling constant of strong interactions, anomalous dimensions
of the nonsinglet operators mentioned above and the standard $\beta$
function, respectively.\\

     The moments of the nonsinglet structure functions have the following
perturbative expansion:

$$
M_{i}(n,Q^{2})=\delta_{i}A_{n}\alpha_{s}^{d_{n}}(Q^{2})\{1+{\alpha_{s}
(Q^{2})\over 4\pi}f^{(2)}_{i,n}+{\alpha_{s}^{2}(Q^{2})\over (4\pi)^{2}}
f^{(3)}_{i,n}+...\},\eqno{(3)}$$
where
$$A_{n}=A_{n}(Q^{2}_{0})\alpha_{s}^{-d_{n}}(Q^{2}_{0})\{1+{\alpha_{s}
(Q^{2}_{0})\over 4\pi}Z^{(2)}_{n} + {\alpha_{s}^{2}(Q^{2}_{0})\over
(4\pi)^{2}}[Z^{(3)}_{n} + {1\over 2}(Z^{(2)}_{n})^{2}]+...\}^{-1}
\eqno{(4)}$$
and  $d_{n},~Z^{(k)}_{n},~f^{(k)}_{n,i}$  are in principle calculable
quantities. The superscript (NS) in (3) and (4) is suppressed.\\

     We remind that the coefficients $Z_{n}^{(k)NS}$ are connected
with the perturbative expansion of
\begin{eqnarray*}
{}~~~exp\{-\int _{\alpha_{s}(Q^{2}_{0})}^{\alpha_{s}(Q^{2})}da{\gamma^{NS}
_{n}(a)\over \beta(a)}\}=({\alpha\over \alpha_{0}})^{d^{NS}_{n}}
\{1 &+& {\alpha-\alpha_{0}\over 4\pi}Z^{(2)NS}_{n}+{\alpha^{2}-\alpha^{2}
_{0}\over (4\pi)^{2}}Z^{(3)NS}_{n}\\ &+&{(\alpha-\alpha_{0})^{2}\over
(4\pi)^{2}}{1\over 2}(Z^{(2)NS}_{n})^{2} + ...\}.~~~~~~~(5)
\end{eqnarray*}

In the RHS of (5) $\alpha=\alpha_{s}(Q^{2})$ and $\alpha_{0}=\alpha_{s}
(Q^{2}_{0}).\\

     In the NLLA (one-loop approximation for the coefficient functions
$C^{NS}_{i,n}$ and two-loop approximation for the functions
$\gamma^{NS}_{n},~\beta$) the moments of the structure functions have the
form

\begin{eqnarray*}
{}~~~~~~M_{i}^{NS}(n,Q^{2}) = \delta_{i}^{NS}A^{NS}_{n}(Q^{2}_{0})\{1 &+&
{\alpha_{s}(Q^{2})\over 4\pi}[Z^{(2)NS}_{n}+c^{(1)NS}_{i,n}]\\
 &-&
{\alpha_{s}(Q^{2}_{0})\over 4\pi}Z^{(2)NS}_{n}\}
\{{\alpha_{s}(Q^{2})\over
\alpha_{s}(Q^{2}_{0})}\}^{d^{NS}_{n}},~~~~~~~~~~(6)
\end{eqnarray*}
where $Z^{(2)NS}_{n},~~d^{NS}_{n}$ and $c^{(1)NS}_{i,n}$ are well
known numbers (e.g. [10]) from the perturbative calculations of the
functions $\gamma^{NS}_{n},~\beta$ and $C^{NS}_{i,n}(1,\alpha_{s})$. In (6)
$~\delta^{NS}_{i}$ are weak or electromagnetic charge factors and $\alpha_
{s}(Q^{2})$ is given in the NLLA

$$
\alpha_{s}(Q^{2})={1\over \beta_{0}ln(Q^{2}/\Lambda^{2})}-{\beta_{1}\over
\beta_{0}^{3}}{lnln(Q^{2}/\Lambda^{2})\over ln^{2}(Q^{2}/
\Lambda^{2})}\eqno{(7)}$$
with      $$b_{0}=11-{2\over 3}N_{f},~~~~~b_{1}=102-{38\over 3}N_{f},$$\\
where $N_{f}$ is the number of flavors.\\

     Let us now rewrite the moments of the nonsinglet structure functions
via the valence quark distributions $V(x,Q^{2})$. Their moments are
defined as

$$V(n,Q^{2}) = \int _{o}^{1}dxx^{n-1}V(x,Q^{2})~.\eqno{(8)}$$

According to the A-definition the $Q^{2}$ evolution of the moments
of the valence quark distributions is given as

$$
V^{(a)}(n,Q^{2}) =
V^{(a)}(n,Q^{2}_{0})L^{(a)}_{n}(Q^{2},Q^{2}_{0})~,\eqno{(9})$$
where
$$
V^{(a)}(n,Q^{2}_{0}) = A^{NS}_{n}(Q^{2}_{0})[1+{\alpha_{s}(Q^{2}_{0})
\over 4\pi}c^{(1)NS}_{2,n}}]\eqno{(10)}$$
and
$$
L^{(a)}_{n}(Q^{2},Q^{2}_{0}) = \{1+{\alpha(Q^{2})-\alpha_{s}(Q^{2}_{0})
\over 4\pi}(Z^{(2)NS}_{n}+c^{(1)NS}_{2,n})\}\{{\alpha_{s}(Q^{2})\over
\alpha_{s}(Q^{2}_{0})}\}^{d^{NS}_{n}}~,\eqno{(11)}$$
i.e. in the case of this definition of the parton distributions the
radiative corrections from the Wilson coefficient functions
$C^{NS}_{2,n}(1,\alpha_{s})$ are included in the evolution of their
moments. In (9) $V^{(a)}(n,Q^{2}_{0})$ are the moments of the valence
quark distributions at some fixed value of $Q^{2}=Q^{2}_{0}$, which
must be determined from the experimental data. The superscript (a) means
that A-definition is used.\\

     Taking into account this definition of the parton distributions the
moments of the structure functions are usually given in the following form:
$$
M^{NS}_{i}(n,Q^{2})=\delta^{NS}_{i}V^{(a)}(n,Q^{2})\{1+{\alpha_{s}(Q^{2})
\over 4\pi}(c^{(1)NS}_{i,n}-c^{(1)NS}_{2,n})\}.\eqno{(12)}$$

It is seen from (12) that in the case $i=2$

$$M^{NS}_{2}(n,Q^{2}) = \delta^{NS}_{2}V^{(a)}(n,Q^{2}),\eqno{(13)}$$
i.e. the naive parton formula for $F_{2}(x,Q^{2})$ is valid in the NLLA too.\\

     According to the B-definition the $Q^{2}$ evolution of the moments of
the valence quark distributions is given as

$$
V^{(b)}(n,Q^{2}) =
V^{(b)}(n,Q^{2}_{0})L^{(b)}_{n}(Q^{2},Q^{2}_{0}),\eqno{(14)}$$
where

$$V^{(b)}(n,Q^{2}_{0})=A^{NS}_{n}(Q^{2}_{0})\eqno{(15)}$$
and
$$
L^{(b)}_{n}(Q^{2},Q^{2}_{0}) =\{1+{\alpha_{s}(Q^{2})-\alpha_{s}(Q^{2}_{0})
\over 4\pi}Z^{(2)NS}_{n}\}\{{\alpha_{s}(Q^{2})\over \alpha_{s}(Q^{2}_{0})
}\}^{d^{NS}_{n}}~.\eqno{(16)$$

Taking into account the B-definition of the parton distributions the
moments of the structure functions are usually given by
$$
M^{NS}_{i}(n,Q^{2}) = \delta^{NS}_{i}V^{(b)}(n,Q^{2})[1+{\alpha_{s}(Q^{2})
\over 4\pi}c^{(1)NS}_{i,n}].\eqno{(17)}$$

    Using (9-11) and (14-16) for A and B-definition of the valence quark
distributions, respectively, one can see that the $Q^{2}$ evolution
equations for the moments of the structure functions (12) and (17) contain
$O(\alpha^{2}_{s})$ terms, which do not attend in (6) and do not belong
to NLLA for the moments.\\

     We shall show now that there exists a representation for the
structure functions written in terms of parton distributions, in which
this weak point is overcome.\\

     Let us use the A-definition (9-11) for the moments of the valence
quark distributions. Then from (6) more precise expressions for the
moments of nonsinglet structure functions in NLLA of QCD can be obtained:
$$
M^{NS}_{i}(n,Q^{2}) = \delta^{NS}_{i}\{V^{(a)}({n},Q^{2})+
{\alpha_{s}(Q^{2})\over 4\pi}(c^{(1)NS}_{i,n}-c^{(1)NS}_{2,n})V(n,Q
^{2})_{"LO"}\},\eqno{(18)}$$
where
$$
V(n,Q^{2})_{"LO"}=V^{(a)}(n,Q^{2}_{0})\{{\alpha_{s}(Q^{2})\over \alpha_{s}
(Q^{2}_{0})}\}^{d^{NS}_{n}},
\eqno{(19)}$$
i.e. $V(n,Q^{2})_{"LO"}$ satisfy the leading order (LLA) evolution
equations for the moments of the valence quark distributions, in which
for $\alpha_{s}(Q^{2})$ the NLLA (7) is taken.\\

     We would like to stress that $O(\alpha^{2}_{s})$ terms are absent in
(18) and there is one-to-one correspondence between the moments of the
structure functions calculated in NLLA in formal QFT approach (Eq. (6)) and
the moments of the same functions in the same approximation when the
latter are expressed in terms of parton distributions (Eq. (18)). The
difference between our representation (18) and that one given in the
literature (Eq. (11)) is in the second term. Instead of
$V^{(a)}(n,Q^{2})_{NLLA}$ we have $V(n,Q^{2})_{"LO"}$.\\

     We present now our formulae for the $Q^{2}$ evolution of the moments
of the nonsinglet structure functions when the next to NLLA
$O(\alpha^{2}_{s})$ corrections are taken into account:

\begin{eqnarray*}
{}~M^{(3)}_{i}(n,Q^{2})=\delta_{i}V^{(a)}(n,Q^{2})_{NNLLA} &+&
\delta_{i}V(n,Q^{2})_{"LO"}\{{\alpha_{s}\over 4\pi}
(c^{(1)}_{i,n}-c^{(1)}_{2,n})\\
&-&{\alpha_{s}\alpha_{so}\over (4\pi)^{2}}(Z^{(2)}_{n}
+c^{(1)}_{2,n})(c^{(1)}_{i,n}-c^{(1)}_{2,n})~~~~~~~~~~~~(20)\\
&+&{\alpha^{2}_{s}\over (4\pi)^{2}}[(c^{(2)}_{i,n}-c^{(2)}_{2,n})+
(c^{(1)}_{i,n}-c^{(1)}_{2,n})Z^{(2)}_{n}]\} ,
\end{eqnarray*}
where the moments of the valence quark distributions $~V^{(a)}(n,Q^{2})
_{NNLLA}~$ evolve according to
\begin{eqnarray*}
{}~~~V^{(a)}(n,Q^{2})_{NNLLA}&=&V^{(a)}(n,Q^{2}_{0})\{1+{\alpha_{s}
-\alpha_{so}\over 4\pi}(Z^{(2)}_{n}+c^{(1)}_{2,n})\\
& &+{\alpha^{2}_{s}-\alpha^{2}_{so}\over (4\pi)^{2}}[Z^{(3)}_{n}+
{1\over 2}(Z^{(2)}_{n})^{2}+c^{(2)}_{2,n}+c^{(1)}_{2,n}Z^{(2)}_{n}]\\
& &+{\alpha_{so}(\alpha_{so}-\alpha_{s})\over (4\pi)^{2}}(Z^{(2)}_{n}+
c^{(1)}_{2,n})^{2}\}\{{\alpha_{s}\over \alpha_{so}}\}
^{d_{n}}~~~~~~~~~~~~~~(21a)
\end{eqnarray*}
and
$$V^{(a)}(n,Q^{2}_{0})=\{1+{\alpha_{so}\over 4\pi}c^{(1)}_{2,n}+
{\alpha_{so}^{2}\over (4\pi)^{2}}c^{(2)}_{2,n}\}A_{n}(Q^{2}_{0}).
\eqno{(21b)}$$
The superscript NS in (20) and (21a,b) is suppressed.
$\alpha_{s}(Q^{2})$ in the above equations is given by the three-loop
approximation of the $\beta$ function. Ixepting a  part of
$Z^{(3)NS}_{n}$ connected with the three-loop approximation of the
anomalous dimensions $\gamma^{NS}_{n}$
all coefficients in (20) and (21) are already known. There is a big
progress now in the calculations of $~\gamma^{(3)NS}_{n}~$ [8]. Note
also that all combinations of the coefficients in front of $~\alpha_{s}$,
$~\alpha_{s}\alpha_{so}~$  and $~\alpha^{2}_{s}~$  do not depend on the
renormalization scheme used to calculate them.\\

   Applying the convolution theorem to (18) we obtain for the nonsinglet
structure functions ${\cal F}^{NS}_{i}(x,Q^{2})$ (NLLA approximation)
$$
{\cal F}^{NS}_{i}(x,Q^{2})=\delta^{NS}_{i}x\{V^{(a)}(x,Q^{2})+{\alpha_{s}
(Q^{2})\over 2\pi}\int _{x}^{1}{dy\over y}[c^{NS}_{i}({x\over y})-c^{NS}_
{2}({x\over y})]V(y,Q^{2})_{"LO"}\},\eqno{(22)}$$
where
$$
c^{NS}_{1}({x\over y}) - c^{NS}_{2}({x\over y})=-{8\over 3}{x\over y},~~~
c^{NS}_{3}({x\over y})-c^{NS}_{2}({x\over y})=-{4\over 3}(1+
{x\over y}).\eqno{(23)}$$
In (22) $V(x,Q^{2})_{"LO"}$ and $V^{(a)}(x,Q^{2})$ obey the first and
second order Lipatov-altarelli-Parisi equations, respectively:
$$
Q^{2}{dV(x,Q^{2})_{"LO"}\over dQ^{2}}={\alpha_{s}(Q^{2})\over 2\pi}\int _
{x}^{1}{dy\over y}P^{(0)}({x\over y})V(y,Q^2)_{"LO"},\eqno{(24a)}$$
$$Q^{2}{dV^{(a)}(x,Q^{2})\over dQ^{2}}={\alpha_{s}(Q^{2})\over 2\pi}
\int _{x}^{1}{dy\over y}[P^{(0)}({x\over y})+{\alpha_{s}(Q^{2})
\over 2\pi}P^{(1)}({x\over y})]V^{(a)}(y,Q^{2}),\eqno{(24b)}$$
$$V(x,Q^{2}_{0})_{"LO"} = V^{(a)}(x,Q^{2}_{0}).\eqno{(24c)}$$

    Notice that $P^{(1)}(x)$ in (24b) is the modified [3] second order
AP kernel, which corresponds to A-definition of the parton distributions.
We would like to mention also that in  [10]  formula similar to (22) has
been obtained for the longitudinal structure function
$~F_{L}=F_{2}-2xF_{1}$. However, in  [10]  instead of $~V(y,Q^{2})_{"LO"}~$
a pure LLA of $V(y,Q^{2})$ has been used. We consider that in the case of
NLLA to use $V(y,Q^{2})_{"LO"}$ is more correct.\\

     Finally we present our expressions for the moments of the singlet
part of the structure functions:
\begin{eqnarray*}
M^{S}_{i}(n,Q^{2})=\delta^{NS}_{i}\{\Sigma^{(a)}(n,Q^{2}) &+& {\alpha_
{s}(Q^{2})\over 4\pi}(c^{(1)S}_{i,n}-c^{(1)S}_{2,n})\Sigma(n,Q^{2})_{"LO"}\\
&+& {\alpha_{s}(Q^{2})\over 4\pi}2N_{f}(c^{(1)G}_{i,n}-c^{(1)G}_{2,n})
G(n,Q^{2})_{"LO"}\}.~~~~(25)
\end{eqnarray*}

Here $G(n,Q^{2})$ are the moments of the gluon distributions and
$\Sigma(n,Q^{2})$ are the moments of the singlet quark distribution

$$
\Sigma(x,Q^{2})= \sum
_{f}[(q_{f}(x,Q^{2})+\bar{q}_{f}(x,Q^{2})].\eqno{(26)}$$

     Applying the convolution theorem to (25) it is a simple task to
obtain the expressions for the singlet part of the structure functions
themselves.\\

     The formulae for the complete electromagnetic and neutrino
(antineutrino) structure functions are listed in the Appendix.\\

     Let us now generelize our formulae for the structure functions for
any order in $\alpha_{s}$ beyond the LLA. For simplicity we present only
the expressions for the nonsinglet structure
functions (A-definition of the parton distributions is used):

\begin{eqnarray*}
{\cal F}^{(n)NS}_{i}(x,Q^{2}) &=& \delta^{NS}_{i}x\{V^{(n)}_{(a)}(x,Q^{2})\\
&& +\sum _{k=1}^{n-1}\sum _{m=0}^{k-1}{\alpha_{s}^{k-m}(Q^{2})\alpha_{s}^
{m}(Q^{2}_{0})\over (2\pi)^{k}}\int _{x}^{1}{dy\over y}a^{(k,m)NS}_{i}
({x\over y})V(y,Q^{2})_{"LO"}\}.\begin{flushright}(27)\end{flushright}
\end{eqnarray*}

     In the case of $O(\alpha^{n-1}_{s})$ corrections to LLA the valence
quark distributions $~V^{(n)}_{(a)}(x,Q^{2})~$ satisfy the $(n)$-order LAP
equation. $V(y,Q^{2})_{"LO"}$ in the above equation obey the first order
LAP equation, in which $\alpha_{s}(Q^{2})$ is given by $n$-loop
approximation of the $\beta$ function. In (27) $\alpha_{s}(Q^{2})$ is taken
in the same approximation.\\

     If B-definition of the parton distributions is used then in (27)
instead of $~V^{(n)}_{(a)}(x,Q^{2})~$ and $~a^{(k,m)NS}_{i}~$,
$~V^{(n)}_{(b)}(x,Q^{2})~$ and $~b^{(k,m)NS}_{i}~$ have to be taken,
respectively.\\

We remind that the valence parton distributions $~V^{(n)}_{(a)}(x,Q^{2})~$
and $~V^{(n)}_{(b)}(x,Q^{2})~$ satisfy different LAP equations. The
coefficient functions $~b^{(k,m)NS}_{i}$ are also different from those
one in the case of A-definition.\\

     Notice that the above representations of the structure functions are
perturbation theory selfconsistent. For any order in $\alpha_{s}$ beyond
the LLA the terms which belong only to this approximation are present in
(27).\\

\begin{flushleft}
{\large {\bf 3. Conclusion}}\\
\end{flushleft}

     We give a new representation of the next to leading corrections to
the nucleon structure functions in terms of parton distributions. The
terms which belong only to this approximation attend in our expressions
for the structure functions. This representation is generelized for the
higher-order QCD corrections beyond NLLA, which we expect to become
noticeable at the current (HERA) and future (LEP*LHC) experiments.
Finally, we consider such a representation of the structure functions to
be useful for more precise determination of the parton distribution from
the deep inelastic data, and especially, in the case of cross-section
data analysis [11].\\

\begin{flushleft}
{\large {\bf Acknowledgments}}
\end{flushleft}
\vspace*{1mm}

   I wish to thank Professor Abdus Salam, the International Atomic
Energy Agency and UNESCO for the hospitality at the International Centre
for Theoretical Physics in Trieste where this work was completed.\\

\begin{flushleft}
{\large {\bf Appendix}}\\
\end{flushleft}
1. A-definition of the parton distributions\\

     a) Elektromagnetic structure functions in NLLA
\begin{eqnarray*}
{\cal F}_{i}^{\mu(e)N}(x,Q^{2})&=&\sum _{f}e^{2}_{q}[q^{(a)}_{f}(x,Q^{2})+
\bar{q}^{(a)}_{f}(x,Q^{2})]_{NLLA}\\
&&+{\alpha_{s}(Q^{2})\over 2\pi}x
\int_ {x}^{1}{dy\over y}\{\sum_{f}e^{2}_{f}[c^{(q)}_{i}({x\over y})
-c^{(q)}_{2}({x\over y})]\\
&~&.[q^{(a)}_{f}(y,Q^{2})+\bar{q}^{(a)}_{f}(y,Q^{2})]_{"LO"}\\
&~&+<e^{2}>2N_{f}[c^{(G)}_{i}({x\over y}) - c^{(G)}_{2}({x\over y})]
G(y,Q^{2})_{"LO"}\},\begin{flushright}(A1)\end{flushright}\\
{}~~~~~~~~~~~~~~~(i=1,2;~~N &=& p,n).
\end{eqnarray*}

    b) Neutrino and antineutrino structure functions in NLLA\\
     We quote the formulae for isoscalar targets.
\begin{eqnarray*}
{}~~~{\cal F}^{\nu(\bar{\nu})}_{i} = x\Sigma^{(a)}(x,Q^{2})_{NLLA} &+& {\alpha
_{s}(Q^{2})\over 2\pi}x\int _{x}^{1}{dy\over y}\{[c^{(q)}_{i}({x\over y})-
c^{(q)}_{2}({x\over y})]\Sigma(y,Q^{2})_{"LO"}\\
&+& [c^{(G)}_{i}({x\over y})-c^{(G)}_{2}({x\over y})]2N_{f}G(y,Q^{2})_
{"LO"}\},~~~~~~~~~(A2)\\
{}~~~~~(i=1,2)
\end{eqnarray*}
\begin{eqnarray*}
{}~F_{3}^{\nu+\bar\nu}(x,Q^{2})&=&\sum _{f}[q^{(a)}_{f}(x,Q^{2}) -\bar{q}^
{(a)}_{f}(x,Q^{2}]_{NLLA}~~~~~~~~~~~~~~~~~~~~~~~~~~~~~~~~~~~~~(A3)\\&~& +
{\alpha_{s}(Q^{2})\over 2\pi}\int _{x}^{1}
{dy\over y}[c^{(q)}_{3}({x\over y})-c^{(q)}_{2}({x\over y})]
\sum _{f}[q_{f}(y,Q^{2}) -
\bar{q}_{f}(y,Q^{2})]_{"LO"}.
\end{eqnarray*}

      Note that we use notations
$$
c_{i}^{(q)}=c^{(1)S}_{i}=c^{(1)NS}_{i},~~~~ c^{(G)}_{i}=c^{(1)G}_{i}.$$
2. B-definition of the parton distributions\\

    a) Electromagnetic structure functions in NLLA

\begin{eqnarray*}
{\cal F}^{\mu(e)N}_{i}(x,Q^{2})&=&\sum _{f}e^{2}_{f}x[q^{(b)}(x,Q^{2})+
\bar{q}^{(b)}_{f}(x,Q^{2})]_{NLLA}\\ &~&+{\alpha_{s}(Q^{2})\over
2\pi}x\int _{x}^{1}{dy\over y}\{c^{q}_{i}({x\over y})\sum
_{f}e^{2}_{f}[q_{f}(y,Q^{2})+\bar{q}_{f}(y,Q^{2})]_{"LO"}\\
&&+<e^{2}>2N_{f}c^{(G)}_{i}({x\over
y})G(y,Q^{2})_{"LO"}\}.\begin{flushright}(A4)\end{flushright}
{}~~~~~~~~~~~(i=1,2;~ N=p,n)
\end{eqnarray*}

 b) Neutrino and antineutrino structure functions in NLLA

\begin{eqnarray*}
{\cal F}^{\nu(\bar{\nu})}(x,Q^{2}) = x\Sigma^{(b)}(x,Q^{2})_{NLLA}&+&
{\alpha_{s}(Q^{2})\over 2\pi}x\int _{x}^{1}{dy\over y}\{c^{(q)}_{i}({x\over
y})\Sigma(y,Q^{2})_{"LO"}\\ &+&
2N_{f}c^{(G)}_{i}({x\over
y})G(y,Q^{2})_{"LO"}\},\begin{flushright}(A5)\end{flushright}
{}~~~~~~~~~~~~~(i=1,2)
\end{eqnarray*}
\begin{eqnarray*}
{}~F^{\nu+\bar{\nu}}_{3}(x,Q^{2}) &=& \sum _{f}[q^{(b)}_{f}(x,Q^{2})-
\bar{q}^{(b)}_{f}(x,Q^{2})]_{NLLA}\\ &~& + {\alpha_{s}(Q^{2})\over 2\pi}
\int _{x}^{1}{dy\over y}c^{(q)}_{3}({x\over y})\{\sum _{f}[q_{f}(y,Q^{2})
-\bar{q}_{f}(y,Q^{2})]_{"LO"}\}.~~~~~~~~(A6)
\end{eqnarray*}

In (A1-A6) the quark $q^{a(b)}_{f}(x,Q^{2})$, antiquark $\bar{q}^{a(b)}
_{f}(x,Q^{2})$ and $\Sigma^{a(b)}(x,Q^{2})$ distributions satisfy
the second order LAP equations. In the case of A-definition of the parton
distributions the usual second order LAP kernels have to be modified (see
[3]). In (A1-A6) the parton distributions $q_{f}(x,Q^{2})_{"LO"},~ \bar
{q}_{f}(x,Q^{2})_{"LO"},~ \Sigma(x,Q^{2})_{"LO"}$ and $G(x,Q^{2})_{"LO"}$
satisfy the first order LAP equations, in which for $\alpha_{s}(Q^{2})$
the NLLA (7) is taken.
\newpage

\begin{flushleft}
{\large {\bf References}}
\end{flushleft}
\vspace*{2mm}
 1. L.N. Lipatov: Yad. Fiz. 20 (1974) 298; G. Altarelli, G. Parisi: Nucl.
 Phys. B126 (1977) 298.\\
 2. G. Altarelli, R.K. Ellis, G. Martinelli: Nucl. Phys. B143 (1978) 521;
 B146 (1979) 544(E); B157 (1979) 461; J. Kubar-Andre, F.E. Paige: Phys.
 Rev. D19 (1979) 221.\\
 3. M. Diemoz, F. Ferroni, E. Longo, G. Martinelli: Z. Phys. C - Particles
 and Fields 39 (1988) 21.\\
 4. L. Baulieu, C. Kounnas: Nucl. Phys. B141 (1978) 423; J. Kodaira,
 T. Uematsu: Nucl. Phys. B141 (1978) 497.\\
 5. W.A. Bardeen, A.J. Buras., D.W. Duke, T. Muta: Phys. Rev. D18 (1978)
 3998. \\
 6. D.I. Kazakov, A.V. Kotikov: Phys. Lett. B291 (1992) 171.\\
 7. E.B. Zijlstra, W.L. van Neerven: Nucl. Phys. B383 (1992) 525;
    S.A. Larin, J.A.M. Vermaseren: Z. Phys. C - Particles and Fields 57
 (1993) 93.\\
 8. S.A. Larin, T. van Ritbergen, J.A.M. Vermaseren: Preprint NIKHEF -
 H/93-29 (1993).\\
 9. K. Wilson: Phys. Rev. 179 (1969) 1499.\\
10. A.J. Buras: Rev. Mod. Phys. 52 (1980) 199.\\
11. S.I. Bilenkaya, D.B. Stamenov: Mod. Phys. Lett. A7 (1992) 2307.\\

\end{document}